\documentclass[nofootinbib,
 reprint,
 amsmath,amssymb,
 aps, onecolumn
]{revtex4-2}

\usepackage{graphicx}% Include figure files
\usepackage{dcolumn}% Align table columns on decimal point
\usepackage{bm}% bold math
 \usepackage{amsmath}
\usepackage{epstopdf}
\usepackage{float}
\usepackage{hyperref}
\usepackage{enumitem}
\usepackage[caption=false]{subfig}
\usepackage{color}
\usepackage[T1]{fontenc}
\usepackage[utf8]{inputenc}
\usepackage[toc,page]{appendix}
\usepackage[usenames,dvipsnames]{xcolor}
\usepackage[normalem]{ulem}
\usepackage{lipsum, babel}

\usepackage{physics}
\usepackage{tensor}
\usepackage{graphicx}
\usepackage{cleveref}

\usepackage{subfiles}%best loaded last in preamble

\newcommand{\lr}[3]{\left#1 #2\right#3}
\newcommand{\tcr}{\textcolor{red}}

\newcommand{\eq}[1]{\begin{align} #1 \end{align}}
\newcommand{\mcal}[1]{\mathcal{#1}}

\newcommand{\be}{\begin{equation}}
\newcommand{\ee}{\end{equation}}
\newcommand{\ba}{\begin{eqnarray}}
\newcommand{\ea}{\end{eqnarray}}

\newcommand{\beq}{\begin{equation}}
\newcommand{\eeq}{\end{equation}}
\newcommand{\beqa}{\begin{eqnarray}}
\newcommand{\eeqa}{\end{eqnarray}}

\begin{document}

\title{Thermodynamics of charged and accelerating black holes
}

\author{Tomáš Hale}%
\email{tomas.hale@utf.mff.cuni.cz}
\affiliation{%
 Institute of Theoretical Physics, Faculty of Mathematics and Physics,
Charles University, V Holešovičkách 2, 180 00 Prague 8, Czech Republic
}%
\author{David Kubizňák}%
\email{david.kubiznak@matfyz.cuni.cz}
\affiliation{%
 Institute of Theoretical Physics, Faculty of Mathematics and Physics,
Charles University, V Holešovičkách 2, 180 00 Prague 8, Czech Republic
}%
\author{Robert B. Mann}
\email{rbmann@uwaterloo.ca}
\affiliation{%
Department of Physics and Astronomy, University of Waterloo, 
Waterloo, Ontario, N2L 3G1, Canada\\
Department of Applied Mathematics, University of Waterloo, Waterloo, Ontario, N2L 3G1, Canada\\
Perimeter Institute  for Theoretical Physics, 31 Caroline Street North, Waterloo, Ontario, N2L 2Y5, Canada
}%
\author{Jana Menšíková}
\email{jana.mensikova@matfyz.cuni.cz}
\affiliation{%
 Institute of Theoretical Physics, Faculty of Mathematics and Physics,
Charles University, V Holešovičkách 2, 180 00 Prague 8, Czech Republic
}%
\author{Jiayue Yang}
\email{j43yang@uwaterloo.ca}
\affiliation{%
Department of Physics and Astronomy, University of Waterloo, 
Waterloo, Ontario, N2L 3G1, Canada\\
Department of Applied Mathematics, University of Waterloo, Waterloo, Ontario, N2L 3G1, Canada\\
Perimeter Institute  for Theoretical Physics, 31 Caroline Street North, Waterloo, Ontario, N2L 2Y5, Canada
}%

%\date{\today}
\date{January 23, 2025}

\begin{abstract}
We reconsider various C-metric
spacetimes describing charged and (slowly)  accelerating AdS black holes in different theories of (nonlinear) electrodynamics and revisit their thermodynamic properties. Focusing first on the Maxwell theory, we find a parametrization of the metric where we can eliminate the nontrivial "normalization" of the boost Killing vector which was crucial for obtaining consistent thermodynamics in previous studies. We also calculate the Euclidean action using (i) the standard holographic renormalization and (ii) the topological renormalization, showing that in the presence of overall cosmic string tension the two do not agree. These results are also extended  
to accelerating black holes in ModMax and RegMax nonlinear electrodynamics. Interestingly, for  the latter the electrostatic potential picks up a modification, 
that remains to be explained, 
but is consistent with the topological renormalization and the (generalized) Hawking-Ross prescription. Our study indicates  
that thermodynamics of charged accelerating black holes is far from being completely understood.
\end{abstract}
\maketitle

\section{Introduction}

Black holes are perhaps the most thoroughly studied class of objects predicted by Einstein's theory of general relativity. They play a critical role in understanding the effects of strong gravity as well as providing a testing ground for predictions based on semiclassical quantum gravity and the full gauge-gravity duality. Special cases of black holes are described as exact solutions of the Einstein field equations. Among these, one of the most perplexing is the solution describing accelerating black holes, known as the C-metric; see e.g.  \cite{Weyl:1917gp,Ehlers:1962zz,Kinnersley:1970zw, Plebanski:1976gy,Podolsky:2002nk, Podolsky:2003gm, Griffiths:2005qp}.

In a standard setup, such a solution describes a pair of  uniformly accelerated black holes with cosmic strings with nontrivial tensions attached to their horizons, causing the acceleration\footnote{Alternatively, one can consider "regular" C-metric spacetimes, where the acceleration is caused by uniform electric, or gravitational fields extending from infinity, e.g. \cite{ernst1976removal, Astorino:2016ybm, Astorino:2021boj}. While such spacetimes are absent of cosmic strings, their asymptotics is strongly deformed. In this paper we focus on the standard case where the acceleration is induced by cosmic strings. We also focus on the "simple" type D C-metrics,  leaving aside more complicated type I solutions  \cite{Astorino:2023ifg, Barrientos:2023dlf}, or solutions with multiple charges \cite{lu2014c}.}.  In this case, the spacetime features an acceleration horizon and loses energy by radiation,  making the equilibrium description questionable (see, however, \cite{Ball:2020vzo, Ball:2021xwt}).  Thus, in order to be able to describe the thermodynamic properties, we restrict ourselves to   (local) AdS asymptotics and the case of  {\em slow acceleration}, in which case only one black hole is present and "hangs" on a string stretched  from conformal infinity \cite{Appels:2016uha}.  This means we keep the acceleration parameter small enough (compared to the AdS scale), so that neither the acceleration horizon nor the second black hole appears. 
Thermodynamics of such spacetimes (including charges and rotation) were resolved in \cite{ Anabalon:2018ydc, Anabalon:2018qfv, Gregory:2019dtq} (see also \cite{Appels:2016uha, Appels:2017xoe} for previous attempts). Here, the %The resultant 
thermodynamic quantities were independently constructed by standard methods and were shown to agree with the Euclidean action calculation,  using the standard holographic renormalization procedure \cite{Emparan:1999pm}.
However, the resulting thermodynamics %procedure 
is hampered by the issue of finding the normalization of the boost Killing vector, necessary for satisfying the first law. While there is a simple way for finding such a normalization in the uncharged nonrotating case, in the case of charged (and/or rotating) black holes this is more problematic, though a solution has been obtained \cite{Anabalon:2018qfv}  (see, however, a discussion in \cite{Kim:2023ncn}). 

In this paper, we  revisit %reconsider 
the formulation of the thermodynamics of four-dimensional charged slowly accelerated AdS black holes.
We 
explore the full parametric freedom of the C-metric solution and find that consistent thermodynamics can be obtained for many forms. In particular, %in Sec.\ref{sec:maxwell} 
we show how we can, for example, eliminate the dependence on the normalization of the boost Killing vector,  set the symmetrized string tension to zero, or make the leading asymptotic term "AdS-like." We demonstrate consequences of such reparametrizations on the expressions for thermodynamic quantities. We also 
%In Sec.\ref{sec:action} we 
perform the standard counterterm calculation of the  action \cite{Emparan:1999pm} in the general parametrization and compare it to the topological renormalization  \cite{Aros:1999id, Aros:1999kt, Miskovic:2009bm,Olea:2005gb}, showing that in the presence of an overall conical deficit the two do not agree.  

Charged accelerated black holes in Maxwell  theory have  
recently been extended to two theories of nonlinear electrodynamics (NLE)  \cite{Sorokin:2021tge}, namely, the modified Maxwell (ModMax) theory \cite{Bandos:2020jsw} and the regularized Maxwell (RegMax) theory \cite{Tahamtan:2020lvq, Kubiznak:2022vft}. 
 In general, theories of NLE provide a framework for
generalizing Maxwell's theory of the classical electromagnetic field by considering the Lagrangian density to be a general function of the two standard electromagnetic invariants in four dimensions, namely: 
\be 
\mcal{L}_{\text{NLE}}=\mcal{L}(\mcal{S},\mcal{P})\,,
\ee 
 where $F=dB$, in terms of the electromagnetic vector potential $B$, and 
\be \label{eq:EM invariants}
\mathcal{S} = \frac{1}{2} F_{\mu\nu} F^{\mu\nu}\,,\quad 
\mathcal{P} = \frac{1}{2} F_{\mu\nu} \left(*F^{\mu\nu}\right)\,,
\ee
 and the Maxwell theory is restored upon considering 
\be\label{LM} 
{\cal L}_{\mbox{\tiny M}}=-\frac{1}{2}{\cal S}\,.
\ee 
When minimally coupled to gravity, this yields the Einstein equations with the following modified 
energy-momentum tensor
\eq{
G_{\mu\nu}+\Lambda g_{\mu\nu}=8\pi T_{\mu\nu}^{\text{NLE}}\,,\quad T_{\text{NLE}}^{\mu\nu}=-\frac{1}{4\pi}\lr{(}{2F^{\mu\sigma}\tensor{F}{^\nu_\sigma}\mcal{L}_{\mcal{S}}+\mcal{P}\mcal{L}_\mcal{P}g^{\mu\nu}-\mcal{L}g^{\mu\nu}}{)}\,,
\label{NLEFE}
}
where $\mcal{L}_\mcal{S}\equiv\pdv{\mcal{L}}{\mcal{S}}$, $\mcal{L}_\mcal{P}\equiv\pdv{\mcal{L}}{\mcal{P}}$ and $\Lambda\equiv-3/\ell^2$ defines the AdS radius $\ell$. The equations for the Faraday tensor $F_{\mu\nu}$ can be made to take on a similar form as the standard Maxwell equations by  introducing the NLE field tensor
\begin{eqnarray}\label{eq:NLE tensor}
D_{\mu\nu} \equiv \frac{\partial{\mathcal{L}}}{\partial{F^{\mu\nu}}}=-2\Bigl({\cal L}_{\cal S} F_{\mu\nu}+{\cal L}_{\cal P} *F_{\mu\nu}\Bigr)\,,
\end{eqnarray}
which in general has a nonlinear dependence on $F_{\mu\nu}$. 
The sourceless NLE equations then read
\begin{eqnarray}\label{eq:NLE equations}
dF=0\,,\quad   d*D=0\,.
\end{eqnarray}
Consequently, we have the following modified prescription for magnetic and electric charges
\be\label{QmQ} 
Q_m=\frac{1}{4\pi}\int_{S^2} F\,,\quad Q=\frac{1}{4\pi} \int_{S^2}*D\,.
\ee 

To date  only two NLE theories are known to admit a C-metric solution: RegMax and ModMax  NLEs. 
{\em ModMax} is the unique NLE that is both invariant under conformal transformations of the metric and possesses full electro-magnetic duality \cite{ModMax}. 
Its Lagrangian density reads
\be
\mathcal{L}_{\mbox{\tiny ModM}}=-\frac{1}{2}\left(\mathcal{S}\cosh{\gamma}-\sqrt{\mathcal{S}^2+{\mathcal{P}^2}}\sinh{\gamma}\right)\,,
\ee
where $\gamma$ is a dimensionless parameter taking on values between $0$ and $\infty$. It reduces to Maxwell case, \eqref{LM},  %$\mathcal{L}_{M}=-\frac{1}{2}\mathcal{S}$ 
for the choice $\gamma =0$, whereas  nonvanishing $\gamma$ implies the
occurrence of a birefringence phenomenon \cite{ModMax,Kosyakov:2020wxv}.
 The C-metric solution in this theory and its thermodynamics in the standard parametrization (defined below) have recently been obtained \cite{Barrientos:2022bzm};  other exact self-gravitating solutions in this theory have subsequently been found  \cite{Flores-Alfonso:2020euz, BallonBordo:2020jtw, Tahamtan:2023tci, Barrientos:2024umq, Bokulic:2025usc}. 

 {\em RegMax} NLE was first derived in \cite{Tahamtan:2020lvq} as the unique NLE permitting radiative solutions in the Robinson-Trautman class of spacetimes.  RegMax is also the only  nontrivial $[{\cal L}={\cal L}({\cal S})]$ NLE admitting "Maxwell-like" slowly rotating charged black holes \cite{Kubiznak:2022vft}. Further study then led to the discovery of the RegMax C-metric solution and other interesting features including extended black hole thermodynamics in \cite{Kubiznak:2023emm}. 
 Studies of the implications for geodesic motion around static RegMax black holes and their optical properties were subsequently carried out  \cite{RegMax:2024}. The corresponding Lagrangian reads 
\eq{
\mathcal{L}_{\mbox{\tiny RegM}}=-2\alpha^4\left(1-3\ln(1-s)+\frac{s^3+3s^2-4s-2}{2(1-s)}\right)\,,\quad s={\sqrt[4]{-\frac{\mathcal{S}}{\alpha^4}}} \in (0,1) \,,
} 
where $\alpha$ is a  dimensionful parameter with dimensions of $[\alpha]=L^{-\frac{1}{2}}$;  in the $\alpha\to\infty$ limit
Maxwell's  theory is restored. This theory leads to a minimally regularized spherically symmetric potential
\eq{
B_{\mbox{\tiny M}}=-\frac{e}{r}\dd t\quad\to\quad B_{\mbox{\tiny RegM}}=-\frac{e}{r+r_0}\dd t\, ,\quad r_0\equiv\sqrt{\abs{e}}/\alpha\,. 
}

In the rest of this paper, we shall first revisit the thermodynamics of charged C-metrics in Maxwell theory (Sec.~\ref{sec:maxwell}) and then turn to the ModMax (Sec.~\ref{modmax}) and RegMax (Sec.~\ref{regmax}) cases.

\section{Charged accelerating black holes in Maxwell electrodynamics} \label{sec:maxwell}

%%%%%%%%%%%%%%%%%%%%%%%%%%%%%%%%%%%%%%%%%%%%%%%%%%%%%%%%%%%%%%%%%%%%%%%%%%%%%%
\subsection{Solution}

The charged C-metric in Maxwell's theory is most easily written in the $x-y$ coordinate system. %Following appendix A in ,
The solution reads \cite{Kubiznak:2023emm} 
\ba\label{eq:Cmetric XY ansatz M}
\dd s^2&=&\frac{1}{H(x,y)^2}\Bigl(-F(y)\dd t^2+\frac{\dd y^2}{F(y)}+\frac{\dd x^2}{G(x)}+G(x)\dd\varphi^2\Bigr)\,,\nonumber\\
B&=&-ey\dd t\,,
\ea
where $H(x,y)=(x+y)$, and 
\eq{
F(y)=&e^2 y^4+\frac{c_1}{6}y^3-\frac{c_2}{2}y^2+c_3 y-c_4+\frac{1}{\ell^2}\,,\nonumber\\
G(x)=&-e^2x^4+\frac{c_1}{6}x^3+\frac{c_2}{2}x^2+c_3 x+c_4\,.
}
Here, $e$ is related to the charge of the black hole, $\ell$ is the cosmological radius, and $\{c_1, c_2, c_3, c_4\}$ are four integration constants. Note that the solution is invariant under a coordinate shift
\be 
x\to x+c\,,\quad y \to y-c\,, 
\ee 
for an arbitrary constant $c$, upon  due redefinition of the integration constants $c_i$. For this reason, only three of the four integration constants $c_i$ are independent--they are related to three physical parameters--the mass, acceleration, and overall conical deficit--of the solution.

In what follows, we shall change coordinates and reparameterize the metric, introducing explicitly the acceleration parameter ${\cal A}$ and the overall conical deficit $K$ as follows,
\be\label{eq:trApp}
y=\frac{1}{r}\,,\quad x\to {\cal A} x\,,\quad {\varphi\to \frac{\varphi}{{\cal A} K}}\,,\quad t\to \frac{t}{\omega}\,, 
\ee
while introducing new metric functions
\be\label{eq:trApp2} 
\Omega=r H\,,\quad f_M=r^2F\,,\quad h_M=\frac{G}{{\cal A}^2}\,.
\ee 
Here, $\omega$ is not a new parameter but rather stands for the parametrization of the timelike Killing vector--it is a function of other parameters to be determined later from thermodynamic considerations.

To proceed further, we demand that the function $h_M$ has roots at $x=\pm 1$ (and is positive in between). This fixes two of the integration constants, for example $c_3$ and $c_4$. Namely, imposing  
\be 
h_M(x=+1)=0=h_M(x=-1)\,,
\ee 
we find  
\be
c_3=-\frac{c_1 \mathcal{A}^2}{6}\,,\quad c_4=e^2 \mathcal{A}^4-\frac{c_2 \mathcal{A}^2}{2}\,.
\ee
Our new metric functions then take on the form
\ba\label{eq:fM Cmetric}
f_M&=&{\frac{r^2}{\ell^2}\Bigl(1-{\cal A}^2\ell^2\bigl[e^2{\cal A}^2-\frac{c_2}{2}\bigr]\Bigr)
-\frac{{\cal A}^2c_1}{6}r-\frac{c_2}{2}+\frac{c_1}{6r}+\frac{e^2}{r^2}}
\,,\nonumber\\
h_M&=&(1-x^2) \left(\mathcal{A}^2 e^2 \left(x^2+1\right)-\frac{\mathcal{A} c_1 x}{6}-\frac{c_2}{2}\right)\,, \label{eq:hM}\\
\Omega &=& 1+\mathcal{A}rx\,, \nonumber
\ea
and the solution reads
\ba\label{eq:line element Maxwell Cmetric}
\dd s^2&=&\frac{1}{\Omega^2}\Bigl(-\frac{f_M \dd t^2}{\omega^2}+\frac{\dd r^2}{f_M}+\frac{r^2 \dd x^2}{h_M}+\frac{r^2 h_M \dd\varphi^2}{K^2}\Bigr)\,,\nonumber\\
B&=&-\frac{e}{r}\frac{\dd t}{\omega},
\ea
where  $\varphi$ is assumed to have periodicity $2\pi$. Moreover, 
for the solution to be well-behaved, we need to further restrict the parameter range so that, for example, $h_M>0$ in the interval $x\in (-1,1)$. This restricts our choice of parameters $c_1$, $c_2$, see e.g. \cite{Abbasvandi:2018vsh} for the analogous discussion.\footnote{In this respect an especially interesting natural choice for standard AdS asymptotics is  $c_2=2\mcal{A}^2 e^2$, upon which the leading asymptotic term in $f_M$ simply reads $r^2/\ell^2$, while at the same time  
\be 
h_M=(1-x^2){\cal A} x\Bigl(\mathcal{A}e^2x-\frac{c_1}{6}\Bigr)\,,
\ee 
which has a new root at $x=0$. Restricting to $c_1<0$ and  the range $x\in (0,1)$, we may still have a black hole spacetime; see the Appendix for more details.
}

In what follows, we shall keep $c_1$ and $c_2$ arbitrary. Note, however, that they only give rise to one more independent physical parameter. We shall employ this freedom when recovering convenient parametrizations of resultant thermodynamic quantities.

%%%%%%%%%%%%%%%%%%%%%%%%%%%%%%%%%%%%
%%%%%%%%%%%%%%%%%%%%%%%%%%%%%%%%%%%
\subsection{Thermodynamic charges and extended first law}
\label{general}
Following the procedure developed in \cite{Appels:2016uha, Appels:2017xoe, Anabalon:2018ydc, Anabalon:2018qfv} (see also \cite{Ferrero:2020twa, Cassani:2021dwa, Boido:2022iye, Kim:2023ncn}), let us now present the  (generalized) thermodynamic charges for the above solution, keeping $c_1$ and $c_2$ as independent parameters. These two are, of course, related to $r_+$, via the horizon defining equation
\be\label{HorizonEq}
f_M(r_+)=0\,, 
\ee
expressing for example
\be\label{c1_eq}
c_1=\frac{r_+^2(3c_2-6{\cal A}^2e^2)-6e^2}{r_+}+\frac{6 r_+^3}{\ell^2({\cal A}^2 r_+^2-1)}\,. 
\ee

Assuming the existence of a black hole in the interval $x\in(-1,1)$, we find the following thermodynamic quantities
\ba \label{TDs_general}
M&=&\frac{c_1}{24 K \omega } \left(\mathcal{A}^2\ell^2\lr{(}{4\mathcal{A}^2e^2-c_2}{)}-2\right)\,,\nonumber\\
T&=&\frac{f_M'(r_+)}{4\pi\omega}=
\frac{c_2 \ell^2 r_+^2 \left(\mathcal{A}^2 r_+^2-1\right)^2-2 e^2 \ell^2 \left(\mathcal{A}^2 r_+^2-1\right)^3+2 r_+^4 \left(\mathcal{A}^2 r_+^2-3\right)}{8 \pi  \ell^2 r_+^3 \omega  \left(\mathcal{A}^2 r_+^2-1\right)}\,,\nonumber\\
S&=&\frac{\mbox{Area}}{4}=
\frac{\pi  r_+^2}{K(1-\mathcal{A}^2 r_+^2)}\,,\\
\phi&=&-B\cdot \xi\big|_{H}=\frac{e}{r_+\omega}\,,\quad 
Q=\frac{1}{4\pi}\int_{S^2}{ *\dd B}=\frac{e}{K}\,, \nonumber
\ea
where the black hole mass was computed using the conformal method \cite{Ashtekar:1999jx}.  This method entails considering a conformal completion of the physical metric $\Bar{g}_{\mu\nu}=\Bar{\Omega}^2g_{\mu\nu}$,  with the Weyl factor
\ba
\bar{\Omega}=\frac{\ell}{r} \Omega\,,
\ea
in order to get rid of the divergence at the boundary. 
The mass can then be viewed as a conserved charge associated with the Killing vector $\xi=\partial_t$. To compute it, we simply need to integrate over the corresponding conserved current associated with the electric part of the Weyl tensor
$\tensor{\bar{C}}{^\nu_{\alpha\mu\beta}}$ of the rescaled metric, namely
\be\label{eq:conserved charge formula}
M=\texttt{Q}(\xi)=\frac{\ell}{8\pi}\lim_{\bar{\Omega}\to 0}\oint\frac{\ell^2}{\bar{\Omega}}N^\alpha N^\beta \tensor{\bar{C}}{^\nu_{\alpha\mu\beta}}\xi_\nu \dd\bar{S}^\mu\,,
\ee
where $N_\alpha$ denotes the normal to the boundary, $N_\alpha=\partial_\alpha \bar{\Omega}$, and the area element $\dd S_\mu$ on the conformal boundary can be expressed as 
\ba\label{eq:area element on conformal boundary}
\dd\bar S_\mu=\delta_\mu^t\lim_{\Bar{\Omega}\to 0} \sqrt{-\Bar{g}}\dd x \dd\varphi=\delta^t_\mu\frac{\ell^2}{K\omega}\dd x \dd\varphi\,,
\ea
where the limit $\bar{\Omega} \to 0$ is obviously equivalent to the condition
$r \to -\frac{1}{\mathcal{A}x}$, i.e., $ \dd r=\frac{\dd x}{\mathcal{A}x^2}$
in the induced metric.

In addition to the above standard thermodynamic quantities, accelerating black holes are characterized by tensions of the cosmic strings attached to the horizon. The acceleration is caused by the difference in tensions of cosmic strings that are attached to its north and south poles. These induce conical deficits $\delta^\pm$ along their axes that are proportional to the corresponding tensions $\mu^\pm$. Namely, we have 
\be\label{eq:string tensions def Maxwell}
8\pi\mu^\pm=\delta^{\pm}=2\pi\Bigl(1-\frac{h_M}{K(1-x^2)}\Bigr)\bigg|_{x=\pm 1}  = \frac{\pi}{3 K}  \left(\pm \mcal{A} c_1+3 c_2-12 \mathcal{A}^2 e^2+6 K\right)\,.
\ee
Rather than $\mu^+$ and $\mu^-$, we will use their symmetric and antisymmetric combinations $(\mu^+ + \mu^-)$ and $(\mu^- - \mu^+)$; the former is usually conveniently simplified and the latter has an instructive physical meaning--it gives the overall force pulling the black hole that causes its acceleration \cite{Gregory:2019dtq}.
Thus, we get
\be\label{eq:mupp Maxwell}
\mu_+\equiv\mu^+ + \mu^- = \frac{-4 \mathcal{A}^2 e^2+c_2+2 K}{4 K}\,,\quad 
\mu_-\equiv\mu^- - \mu^+ = -\frac{\mathcal{A} c_1}{12 K}\,.
\ee
Following \cite{Appels:2017xoe}, we shall call the corresponding conjugate potentials $\lambda_+$ and $\lambda_-$  "thermodynamic lengths."
We also identify the negative cosmological constant with positive thermodynamic pressure \cite{Kastor:2009, Kubiznak:2016qmn}
\be\label{eq:pressure def}
P=\frac{3}{8\pi \ell^2}\,,
\ee 
and call the conjugate quantity the thermodynamic black hole volume $V$. 

It is now easy to verify  that the above thermodynamic quantities obey the extended first law, together with the corresponding generalized Smarr relation:\footnote{The Smarr relation can be derived from the first law by Euler's theorem of homogeneous functions, e.g. {\cite{Kastor:2009},  
which states that 
\be 
f(\alpha^p x, \alpha^q y,\dots)=\alpha^r f(x,y,\dots) \quad \Rightarrow \quad r f(x,y,\dots)=p \Bigl(\frac{\partial f}{\partial x}\Bigr)x+q\Bigl(\frac{\partial f}{\partial y}\Bigr)y+\dots\,.
\ee 
For the black hole mass $M$,  this essentially boils down to a dimensional analysis of its proper variables. In our case, we have $M=M(S, Q, \mu_+, \mu_-, P)$, and taking into account their following scaling: $(M, S,  Q, \mu_+, \mu_-, P) \sim (L^1, L^2, L^1, L^0, L^0,  L^{-2})$ in length dimensions, which follows from $\ell, r_+, e, c_1 \sim L^1$, $\omega, K, c_2 \sim L^0$ and $\mathcal{A} \sim L^{-1}$, gives 
\be 
1\times M=2\times S\Bigl(\frac{\partial M}{\partial S}\Bigr)+1\times Q\Bigl(\frac{\partial M}{\partial Q}\Bigr)+0\times \mu_+\Bigl(\frac{\partial M}{\partial \mu_+}\Bigr)+0\times \mu_-\Bigl(\frac{\partial M}{\partial \mu_-}\Bigr)-2\times P\Bigl(\frac{\partial M}{\partial P}\Bigr)\,,
\ee 
which, upon using 
\eqref{eq:first law Maxwell1} yields the Smarr relation
\eqref{eq:first law Maxwell1b}.}
}
%the second relation therein. 
%thereby yielding   \eqref{eq:first law Maxwell1}. Note also that for the conjugate quantities, one has $(T, V, \phi, \lambda_+, \lambda_-) \sim %(L^{-1}, L^3, L^0, L^1, L^1)$.}
\ba
\delta M&=&T\delta S+\phi \delta Q+\lambda_+ \delta\mu_+ + \lambda_- \delta\mu_- +V\delta P\,,\label{eq:first law Maxwell1}\\
M&=&2(TS-PV)+\phi Q\,,\label{eq:first law Maxwell1b}
\ea
provided we choose\footnote{{The above expression for $\omega$ is valid provided we take $c_2<2 {\cal A}^2e^2$. When $c_2\geq 2 {\cal A}^2e^2$, another root for $h_M$ appears in the interval $x\in(-1,1)$ and the entire thermodynamics has to be redone from scratch; see the Appendix for the corresponding discussion of the case $c_2=2 {\cal A}^2e^2$. Specifically, this implies that it does not make sense to plug $c_2=4 {\cal A}^2e^2$ in the above formula for which $\omega$ would apparently vanish.}} 
\be\label{eq:omega Maxwell generic solution 1}
\omega=\frac{1}{2}\sqrt{\left(4 \mathcal{A}^2 e^2-c_2\right) \left(2-4 \mathcal{A}^4 e^2 \ell^2+\mathcal{A}^2 c_2 \ell^2\right)}\,,
\ee
together with the following thermodynamic lengths
\ba
\lambda_+&=&\left(\frac{\partial M}{\partial \mu_+}\right)_{S,Q,P,\mu_-}={\frac{c_2 \ell^2 r_+^2 \left(\mathcal{A}^2 r_+^2+1\right)-2 e^2 \ell^2 \left(\mathcal{A}^4 r_+^4+4 \mathcal{A}^2 r_+^2-1\right)+2 r_+^4}{2\ell^2 r_+ \omega \left(\mathcal{A}^2 r_+^2-1\right) \left(c_2-4 \mathcal{A}^2 e^2\right)}}\,, \label{eq:string potentials def}\nonumber\\
\lambda_-&=&\left(\frac{\partial M}{\partial \mu_-}\right)_{S,Q,P,\mu_+}={\frac{\mathcal{A} \left(-4 \mathcal{A}^4 e^2 \ell^2 r_+^2+c_2 \ell^2 \left(\mathcal{A}^2 r_+^2-1\right)+4 \mathcal{A}^2 e^2 \ell^2+2 r_+^2\right)}{2 \omega\left(\mathcal{A}^2 r_+^2-1\right)}}\,,
\ea
and volume
\ba\label{eq:volume def}
V&=&\left(\frac{\partial M}{\partial P}\right)_{S,Q,\mu_-,\mu_+}\nonumber\\
&=&\pi\frac{\mathcal{A}^4 \left(c_1 \ell^4 \left(c_2 r_+^2+4 e^2\right)+24 e^2 \ell^2 r_+^3\right)-\mathcal{A}^2 \ell^2 \left(c_1 c_2 \ell^2-4 c_1 r_+^2+12 c_2 r_+^3-24 e^2 r_+\right)-4 \mathcal{A}^6 c_1 e^2 \ell^4 r_+^2-24 r_+^3}{18 K \omega\left(\mathcal{A}^2 r_+^2-1\right)}\,.
\ea   

Before we proceed further, let us briefly comment on cohomogeneity of the obtained first law \eqref{eq:first law Maxwell1}. We started with the parametrization \eqref{eq:Cmetric XY ansatz M} characterized by six parameters 
$\{e,\ell, c_1, c_2, c_3, c_4\}$ (of which only five are physical). Setting the roots of $h_M$ for convenience to $\pm1$, we have eliminated two constants $c_3$ and $c_4$, and traded them for two new parameters $K$ and ${\cal A}$. We have also introduced a shorthand $\omega$--a function of other parameters.  Furthermore, by employing the horizon equation \eqref{HorizonEq}, we may exchange the mass parameter $c_1$ for $r_+$. The resultant parameters to vary are thus
\be 
\{r_+, {\cal A}, K, e, \ell, c_2\}\,,
\ee 
while $\omega$ is expressed in terms of these via \eqref{eq:omega Maxwell generic solution 1}. The obtained  first law \eqref{eq:first law Maxwell1} with five terms on the right-hand side is thus of full cohomogeneity. However, contrary to the first law formulated in \cite{Anabalon:2018qfv} we now allow for six independent variations. Since there are only five conjugate pairs of physical quantities characterizing the properties of the black hole, we are allowed to eliminate one of the parameters above, say $c_2$. This freedom allows one to simplify the above thermodynamic quantities, as we shall now demonstrate. Note also 
%In what follows we shall $c_2$ here is truly not a  in fact it admits 
that the standard first law is recovered by restricting the variations to obey $\delta\mu_+=0=\delta\mu_-=\delta P$.

\subsection{Useful parametrizations}

Let us now exploit the freedom in the choice of $c_2$ to obtain simplified expressions for our thermodynamic quantities. For convenience, we shall also conveniently rename the mass parameter $c_1$.

\subsubsection{Standard parametrization}

First, as a sanity check, let us make the following standard choice (see e.g. \cite{Appels:2016uha, Anabalon:2018qfv}):
\be\label{eq:standard c1 Maxwell}
c_1=-12m\,,\quad 
c_2=2 \mathcal{A}^2 e^2-2\,. 
\ee
It is easy to verify that upon such a choice, we recover the metric functions and thermodynamic charges presented  %read:
in \cite{Anabalon:2018qfv}, namely
\ba
    f_M&=&\left(1-\mathcal{A}^2 r^2\right) \left(1-\frac{2 m}{r}+\frac{e^2}{r^2}\right)+\frac{r^2}{\ell^2}\,,\quad 
    h_M=\left(1-x^2\right)\left(1+2 \mathcal{A} m x+\mathcal{A}^2 e^2 x^2\right)\,,\nonumber
\label{eq:f Maxwell stndard para}\\
    M&=&\frac{m \left(1-\mathcal{A}^4 e^2 \ell^2-\mathcal{A}^2 \ell^2\right)}{K \omega }\,,\nonumber\label{eq:mass Maxwell stndard para}\\ 
    T&=&\frac{e^2 \ell^2 \left(\mathcal{A}^2 r_+^2-1\right)^2-r_+^2 \left(\ell^2 \left(\mathcal{A}^2 r_+^2-1\right)^2-\mathcal{A}^2 r_+^4+3 r_+^2\right)}{4 \pi  \ell^2 r_+^3 \omega  \left(\mathcal{A}^2 r_+^2-1\right)}\,,\quad S=\frac{\pi  r_+^2}{K(1-\mathcal{A}^2 r_+^2)}\,,\nonumber\label{eq:temp Maxwell stndard para}\\ 
    \mu_+&=&\frac{-\mathcal{A}^2 e^2+K-1}{2 K}\,,\quad 
    \lambda_+=\frac{e^2 \ell^2 \left(3 \mathcal{A}^2 r_+^2-1\right)+\ell^2 \left(\mathcal{A}^2 r_+^4+r_+^2\right)-r_+^4}{2 \ell^2 r_+ \omega  \left(\mathcal{A}^2 e^2+1\right) \left(\mathcal{A}^2 r_+^2-1\right)}\,,\label{eq:lamp Maxwell stndard para}\\
    \mu_-&=&\frac{\mathcal{A} m}{K}\,,\quad 
    \lambda_-=\frac{\mathcal{A} \left(r_+^2-\ell^2 \left(\mathcal{A}^2 e^2+1\right) \left(\mathcal{A}^2 r_+^2-1\right)\right)}{\omega  \left(\mathcal{A}^2 r_+^2-1\right)}\,,\label{eq:lamm Maxwell stndard para}\nonumber\\
    V&=&\frac{2 \pi  \left\{ \mathcal{A}^8 e^2 \ell^4 r_+^4 \left(e^2+r_+^2\right)-\mathcal{A}^6 \left(2 e^4 \ell^4 r_+^2+e^2 \ell^2 r_+^4 \left(\ell^2+r_+^2\right)-\ell^4 r_+^6\right)+\right.}{3 K r_+ \omega  \left(\mathcal{A}^2 r_+^2-1\right)^2}\nonumber\\
    &&\left.\scalebox{0.9}{$\mathcal{A}^4 \ell^2 \left(e^4 \ell^2+e^2 \left(r_+^4-\ell^2 r_+^2\right)-r_+^4 \left(2 \ell^2+r_+^2\right)\right)+\mathcal{A}^2 \ell^2 \left(e^2 \ell^2+\ell^2 r_+^2+r_+^4\right)+2 r_+^4$}\right\} \,,\label{eq:volume Maxwell stndard para}\nonumber\\
    \omega &=& \sqrt{\left(\mathcal{A}^2 e^2+1\right) \left(1-\mathcal{A}^4 e^2 \ell^2-\mathcal{A}^2 \ell^2\right)}\,.\nonumber
    \ea 
The corresponding thermodynamic behavior, including a novel phenomenon of "snapping swallow tails," was analyzed in \cite{Abbasvandi:2018vsh} (see also \cite{Abbasvandi:2019vfz} for extensions to the rotating case).

\subsubsection{Normalized Killing vector}
Perhaps more interesting is to choose the parameters $c_1$ and $c_2$ to eliminate $\omega$ and to  simplify the physical mass, namely to set
\be 
\omega=1\,,\quad M=\frac{m}{K}\,.
\ee
The former is possible, as 
the asymptotics of the charged C-metric is affected by the acceleration, cf. \eqref{eq:fM Cmetric}. Thus, by reparametrizing $c_2$, one may affect the effective normalization of the boost Killing vector $\xi$. Namely, the above is achieved by the following choice of the two constants,
\be
c_1=-\frac{24 m}{\beta_+}\,,\quad  
c_2=\frac{4 \mathcal{A}^4 e^2 \ell^2+\beta_-}{\mathcal{A}^2 \ell^2}\,,\quad \beta_\pm=\sqrt{1-4 \mathcal{A}^2 \ell^2}\pm 1\,,
\ee 
upon which the metric functions and the remaining thermodynamic functions are  
\ba
f_M&=&\frac{e^2 \left(\mathcal{A}^2 r^2-1\right)^2}{r^2}+\frac{2 \left(\mathcal{A}^2 r^2-1\right) (2 m-r)}{r \beta_+}+\frac{r^2}{\ell^2}\,,\quad 
h_M=\frac{(1-x^2)\left(\mathcal{A}^2 e^2 \left(x^2-1\right) \beta_++4 \mathcal{A} m x+2\right)}{\beta_+}\,,\nonumber\\
T&=&\frac{-2 e^2 \ell^2 \left(\mathcal{A}^2 r_+^2-1\right)^3+2 \mathcal{A}^2 r_+^6+\frac{r_+^2}{\mathcal{A}^2} \left(\mathcal{A}^2 r_+^2-1\right)^2 \left(4 \mathcal{A}^4 e^2 \ell^2+\beta_-\right)-6 r_+^4}{8 \pi  \ell^2 r_+^3 \left(\mathcal{A}^2 r_+^2-1\right)}\,,\quad S=\frac{\pi  r_+^2}{K(1-\mathcal{A}^2 r_+^2)}\,,\nonumber\\ 
\mu_+&=&\frac{2 \mathcal{A}^2 K \ell^2+\beta_-}{4 \mathcal{A}^2 K \ell^2}\,,\quad 
\lambda_+=-\frac{e^2 \ell^2\beta_+ \left(\mathcal{A}^2 r_+^2-1\right)^2+r_+^4 \beta_+-2 \ell^2 \left(\mathcal{A}^2 r_+^4+r_+^2\right)}{4 \ell^2 r_+ \left(\mathcal{A}^2 r_+^2-1\right)}\,,\\
\mu_-&=&\frac{2 \mathcal{A} m}{K\beta_+}\,,\quad 
\lambda_-=\frac{\mathcal{A}^2 r_+^2\beta_+ -\beta_-}{2 \mathcal{A}^3 r_+^2-2 \mathcal{A}}\,,\nonumber\\
V&=&\frac{\pi  \left[ \mathcal{A}^2 \left(r_+^4 \left(\beta_++2\right)-e^2 \ell^2\beta_- -2 \ell^2 r_+^2\right)+\mathcal{A}^8 e^2 \ell^2 r_+^6 \beta_--r_+^2 \beta_--\mathcal{A}^6 \ell^2 r_+^4 \left(3 e^2 \beta_-+2 r_+^2\right)+\mathcal{A}^4 \ell^2 r_+^2 \left(3 e^2\beta_- +4 r_+^2\right)\right]}{3 \mathcal{A}^2 K r_+ \left(\mathcal{A}^2 r_+^2-1\right)^2}\,.\nonumber
\ea 
Another interesting choice, $c_2=2{\cal A}^2e^2$, follows the same spirit but must be treated separately; we  discuss this in the Appendix.

\subsubsection{Vanishing overall tension}\label{2C3}

Of course many other useful parametrizations are possible. To present a final example, we set
\be
c_1=-12m\,,\quad  
c_2=4 \mathcal{A}^2 e^2-2 K\,,
\ee 
yielding 

\be \label{mup=0}
\mu_+=0\,,\quad \mu_-=\frac{\mathcal{A}m}{K}\,.
\ee
While this would seem to be a mere parameter choice for $c_2$, this is not the case. 
It corresponds to a physical choice of the distribution of the strings, as is obvious from \eqref{mup=0}. With this choice we have eliminated one physical parameter and have $\mu^+=-\mu^-$. 
Note that in this case one of the strings is actually a strut with negative tension.

The metric functions and other thermodynamic quantities are then
\ba
f_M&=&\frac{\left(\mathcal{A}^2 r^2-1\right) \left(e^2 \left(\mathcal{A}^2 r^2-1\right)+r (2 m-K r)\right)}{r^2}+\frac{r^2}{\ell^2}\,,\quad 
h_M=(1-x) (x+1) \left(\mathcal{A}^2 e^2 \left(x^2-1\right)+2 \mathcal{A} m x+K\right)\,,\nonumber\\
T&=&\frac{e^2 \ell^2 \left(\mathcal{A}^2 r_+^2-1\right)^2 \left(\mathcal{A}^2 r_+^2+1\right)-K \ell^2 r_+^2 \left(\mathcal{A}^2 r_+^2-1\right)^2+r_+^4 \left(\mathcal{A}^2 r_+^2-3\right)}{4 \pi  \ell^2 r_+^3 \left(\mathcal{A}^2 r_+^2-1\right) \omega}\,,\quad S=\frac{\pi  r_+^2}{K(1-\mathcal{A}^2 r_+^2)}\,,\nonumber\\ 
\mu_-&=&\frac{\mathcal{A}m}{K}\,,\quad 
\lambda_-=\frac{\mathcal{A} \left(K \ell^2 \left(1-\mathcal{A}^2 r_+^2\right)+r_+^2\right)}{\left(\mathcal{A}^2 r_+^2-1\right) \omega}\,,\nonumber\\
M&=&\frac{m \left(1-\mathcal{A}^2 K \ell^2\right)}{K \omega }=\frac{m\omega}{K^2}\,,\nonumber\\
V&=&\frac{2 \pi  \left(\mathcal{A}^6 K \ell^4 r_+^4 \left(3 e^2+K r_+^2\right)-\mathcal{A}^8 e^2 K \ell^4 r_+^6-\mathcal{A}^4 K \ell^2 r_+^2 \left(3 e^2 \ell^2+2 K \ell^2 r_+^2+r_+^4\right)+\mathcal{A}^2 K \ell^2 \left(e^2 \ell^2+K \ell^2 r_+^2+r_+^4\right)+2 r_+^4\right)}{3 K r_+ \left(\mathcal{A}^2 r_+^2-1\right)^2 \omega}\,.\nonumber\\
\omega &=&\sqrt{K-\mathcal{A}^2 K^2 \ell^2}\,.
\ea 
 Note that in this case we still have one redundant parameter whose value can be fixed to further simplify the thermodynamics.

%%%%%%%%%%%%%%%%%%%%%%%%%%%%%%%%%
%%%%%%%%%%%%%%%%%%%%%%%%%%%%%%%%%%%
\subsection{Action calculation}\label{sec:action}

Full knowledge of a thermodynamic system can be acquired by knowledge of the appropriate free energy $F$. This can be calculated in the semiclassical approximation using
\eq{
F=-\frac{1}{\beta}\log Z\approx\frac{S_E(g_c)}{\beta}\,,
}
where $Z$ is the gravitational partition function, and $S_E$ the Euclidean action evaluated on a classical solution $g_c$ (in our case the charged Euclideanised C-metric) with periodicity $\beta=\frac{1}{T}$ of the Wick-rotated time coordinate.
The standard approach to calculate  such an action is to use the York-Gibbons-Hawking prescription, supplemented by the AdS counterterms, namely \cite{Emparan:1999pm}  
\eq{\label{SEstandard}
S_E=\frac{1}{16\pi G}\int_M\dd^4 x\sqrt{g}\lr{(}{R+\frac{6}{\ell^2}+\mathcal{L}_M}{)}+{\frac{1}{8\pi G}}\int_{\partial M}\dd^3 x\sqrt{h}\mathcal{K}-\frac{1}{8\pi G}\int_{\partial M}\dd^3 x\sqrt{h}\Bigl(\frac{2}{\ell}+\frac{\ell}{2}\mathcal{R}(h)\Bigr)\,,
}
where $h$ is the induced metric on the boundary $\partial M$, with a unit normal $n_\mu$,  and $\mathcal{K}=\nabla_\alpha n^\alpha$ and $\mathcal{R}(h)$ are the extrinsic and intrinsic boundary curvatures, respectively. The first two terms in this action yield the standard Dirichlet boundary value problem and the AdS counterterms (third term) guarantee the overall finiteness.  

For a general parametrization discussed in Sec.~\ref{general},  
the resulting free energy reads
\eq{\label{eq: free energy general c1c2}
F=-\frac{c_1 \left(1+\mathcal{A}^2 c_2 \ell^2-4 \mathcal{A}^4 \ell^2 e^2\right)}{24 K \omega}-\frac{r_+^3}{2 K\omega\ell^2 \left(\mathcal{A}^2 r_+^2-1\right)^2}-\frac{e^2}{2 r_+ K \omega}\,,
}
generalizing the results in \cite{Anabalon:2018qfv}. It is easy to verify that (given the thermodynamic quantities derived above) it satisfies the  relation
\eq{
F=F(T,\phi, \mu_+, \mu_-, P)=M-TS-\phi Q \label{freeen}
}
corresponding to a fixed potential ensemble.

Moreover, one can use the above expression to derive the corresponding conjugate quantities, from the knowledge of thermodynamic potentials $\{T,\phi, \mu_+,\mu_-, P\}$, namely
\ba\label{SSS}
S&=&-\Bigl(\frac{\partial F}{\partial T}\Bigr)_{\phi,  \mu_+,\mu_-, P}\,,\quad 
Q=-\Bigl(\frac{\partial F}{\partial \phi}\Bigr)_{T,  \mu_+,\mu_-, P}\,,\nonumber\\
V&=&\Bigl(\frac{\partial F}{\partial P}\Bigr)_{T, \phi,  \mu_+,\mu_-}\,,\quad
\lambda_+=\Bigl(\frac{\partial F}{\partial \mu_+}\Bigr)_{T, \phi, \mu_-, P}\,,\quad
\lambda_-=\Bigl(\frac{\partial F}{\partial \mu_-}\Bigr)_{T, \phi, \mu_+, P}\,.
\ea
Interestingly, since $S$ is given by the Bekenstein area law, \eqref{TDs_general}, the first relation can be used to derive the unknown factor $\omega$. Namely, one first eliminates $c_1$, by employing the horizon equation \eqref{c1_eq}. Then one calculates the differentials in terms of the parameters $\{{\cal A},K, r_+, e, \ell, c_2\}$, and employs the entropy formula 
\eqref{SSS}, imposing $0=\delta\phi=\delta \mu_+=\delta \mu_-=\delta P$. The latter represents a systems of linear equations for variations of the parameters. The entropy $S$ can thus be written in terms of two (unspecified) variations, say $\{\delta e, \delta c_2\}$. Treating these as independent and comparing the result with  \eqref{TDs_general} then yields two partial differential equations for $\omega=\omega(e, {\cal A}, \ell, c_2)$, whose integration  yields precisely 
\be
\omega\propto  \sqrt{\left(4 \mathcal{A}^2 e^2-c_2\right) \left(2-4 \mathcal{A}^4 e^2 \ell^2+\mathcal{A}^2 c_2 \ell^2\right)}\,,
\label{omint}
\ee
up to a function of the parameter $\ell$.  A dimensional argument then implies that the proportionality factor in 
\eqref{omint} is just a constant. We have thus recovered  
\eqref{eq:omega Maxwell generic solution 1}, up to a constant coefficient (which can be recovered by considering limiting cases of vanishing acceleration). The implications for the phase behavior of the different choices of $c_1$ and $c_2$ is a topic we leave for future study.

\subsection{Action with topological renormalization}
\label{sectopo}

{There  exists an alternative for calculating the  Euclidean action, known as  topological renormalization \cite{Aros:1999id, Aros:1999kt, Miskovic:2009bm,Olea:2005gb} (see e.g. \cite{Ciambelli:2020qny, Corral:2024lva} for its applications to "NUTty spacetimes"). It is instructive to see how this procedure works for highly nontrivial accelerated black hole spacetimes that are penetrated by conical singularities.
} 

\subsubsection{Basic idea}

The idea is to consider
a bulk integral of a topological Gauss-Bonnet term:
\be  
\mathcal{G}=R_{\mu\nu\alpha\beta}R^{\mu\nu\alpha\beta}-4R_{\mu\nu}R^{\mu\nu}+R^2\,. 
\ee
On a four-dimensional manifold without boundaries, the Euler theorem states that $\int_M\dd^4 x \mathcal{G}=32\pi^2\chi(M)$, where $\chi(M)$ is the Euler characteristic
%which is equal to 2 for our spacetime 
\cite{Gibbons:1979xm}. 
%\tco{\bf Are we sure about that?} \tcg{\bf See the reference, under (4.6), which applies without acceleration}. 
On a manifold with boundary we get a correction \cite{Eguchi:1980jx} 
\eq{\label{eq:Chern-Gauss-Bonnet}
\int_M \dd^4 x\sqrt{g} \,\mathcal{G}=32\pi^2\chi(M)+\int_{\partial M}\dd^3 x B_3\,,
}
where $B_3$ is a boundary term--the second Chern form. 
It is this term that can be used to replace the standard boundary terms in \eqref{SEstandard}. Namely, 
one seeks $S_T$ in the form
\eq{
S_T=\frac{1}{16\pi G}\int_M\dd^4 x\sqrt{g}\lr{(}{R+\frac{6}{\ell^2}+\mathcal{L}_M}{)}+\frac{\mathcal{C}}{16\pi G}
\int_{\partial M}\dd^3 x B_3\,.
%\lr{(}{\int_M\dd^4 x\sqrt{g}\, \mathcal{G}-32\pi^2\chi(M)}{)}\,,
}
Evaluating the above action for the spherical Schwarzschild-AdS spacetime, for which $\chi(M)=2$ \cite{Gibbons:1979xm},  it turns out that choosing $\mathcal{C}=\ell^2/4$ automatically yields a finite action (cancelling the divergences coming from the bulk integrals). Moreover, in this case the value of the on-shell $S_T$ agrees with that of the standard action \eqref{SEstandard}. Thus, we have the following alternative prescription for evaluating the on-shell action:
\be \label{ST}
S_T=\frac{1}{16\pi G}\int_M\dd^4 x\sqrt{g}\Bigl(R+\frac{6}{\ell^2}+\mathcal{L}_M+\frac{\ell^2}{4}{\cal G}\Bigr)-S_0\,,
\ee 
where $S_0$ is a constant.
%related to the Euler characteristic of the manifold. 
In particular, for Schwarzschild-AdS we find 
\be \label{S0_0}
S_0=-\frac{\pi \ell^2}{2}\chi(M)=-\pi\ell^2\,,
\ee
upon using $\chi(M)=2$.  The same prescription, with $S_0$ given by \eqref{S0_0}, also works for example for Kerr-AdS black holes, in accordance with Eq.~\eqref{eq:Chern-Gauss-Bonnet}.

\subsubsection{Topological renormalization for accelerating black holes}

Let us now apply topological renormalization to charged  accelerating black holes. Using naively the prescription \eqref{ST} with $S_0$  given by \eqref{S0_0} yields  
\be
S_E=S_T+2\mu_+S_0\,,
\ee
where $\mu_+$ is given by \eqref{eq:mupp Maxwell}. 
Clearly unless $\mu_+=0$ the two actions no longer agree.\footnote{Interestingly, $\mu_+=0$ is a particular  case of a condition $\mu_+=\mbox{const.}$,
derived in \cite{Kim:2023ncn}, where it was argued that such a condition is sufficient for the standard York-Gibbons-Hawking action to yield a well-posed variation principle (with properly vanishing boundary integrals). Moreover, such a condition can be achieved by exploiting the freedom in the parametrization of the solution, see Sec.~\ref{2C3}.
Note also that requiring $\mu_+=0$ does not mean that the black hole cannot accelerate--it only specifies that overall deficit has to be zero.
}

 If we take the above action $S_T$ seriously, we have the following free energy
\be \label{FFT}
F_T=\frac{S_T}{\beta}=F-2\mu_+S_0 T=M-\phi Q-T(S+2\mu_+S_0)\,,
\ee 
which would seem to predict a modified entropy $\tilde S=S+2\mu_+S_0$
 keeping all the potentials as before. This seems a bit analogous to a contribution of Misner strings to Noether entropy as, for example, proposed in \cite{Mann:1999pc}.  Interestingly, the modified entropy is consistent with the first law upon modifying other conjugate quantities, such as $V$ and $\lambda_+$.

 Alternatively, in accordance with standard thermodynamics, let us believe that topological renormalization should yield $S_E$ even in the case of accelerated black holes and for any $\mu_+$. This can be achieved by 
replacing $S_0$ in \eqref{S0_0} by $S_0(1-2\mu_+)$. Formally, this  amounts to replacing in \eqref{eq:Chern-Gauss-Bonnet}
\eq{
\chi(M)=2\quad \rightarrow\quad  {\chi}(M)+\mbox{const}.= 2-4\mu_+\,,}
where the additional constant (const.) takes into account the cosmic string contributions (presumably encoded in additional boundary terms).\footnote{This is in a way similar to a procedure developed in \cite{Cassani:2021dwa}. There, the idea was to consider a quantization of the conical deficits $\delta^{\pm}=2\pi\lr{(}{1-1/n_{\pm}}{)}_{x=\pm 1}$, where $n_\pm$ are coprime positive integers. Then the surfaces of constant $t,r$ become the weighted projective space orbifold $\mathbb{WCP}_{[n_-,n_+]}^1$ also known as a spindle. The orbifold can be assigned a generalized notion of $\chi(M)$: the orbifold Euler characteristic, with rational number values
\eq{
\tilde{\chi}\lr{(}{\mathbb{WCP}_{[n_-,n_+]}^1}{)}=\frac{1}{n_-}+\frac{1}{n_+}=2-4\mu_+\,.
}
In this way, \eqref{eq:Chern-Gauss-Bonnet} retains its form without additional boundary terms and standard thermodynamics from $S_E$ is recovered.
}

\section{ModMax C-metric} \label{modmax}

  Equipped with the detailed knowledge of the Maxwell case, let us now turn toward accelerating black holes in NLE. Let us start with the charged accelerating black holes in the ModMax nonlinear electrodynamics \cite{Bandos:2020jsw} constructed in \cite{Barrientos:2022bzm}. 

\subsection{Solution}

Since on static solutions the ModMax theory  reduces to the Maxwell one, similar to \cite{Barrientos:2022bzm}, we need to consider the presence of magnetic charge. In an $x-y$ coordinate system the solution then reads
\ba\label{eq:Cmetric XY ansatz}
\dd s^2&=&\frac{1}{H(x,y)^2}\Bigl(-F(y)\dd t^2+\frac{\dd y^2}{F(y)}+\frac{\dd x^2}{G(x)}+G(x)\dd\varphi^2\Bigr)\,,\nonumber\\
B&=&-ey \exp(-\gamma)\dd t+gx \dd\varphi\,,
\ea
where $H(x,y)=(x+y)$, and 
\eq{
F(y)=&z^2 y^4+\frac{c_1}{6}y^3-\frac{c_2}{2}y^2+c_3 y-c_4+\frac{1}{\ell^2}\,,\nonumber\\
G(x)=&-z^2x^4+\frac{c_1}{6}x^3+\frac{c_2}{2}x^2+c_3 x+c_4\,,\quad z^2=(e^2+g^2)\exp(-\gamma)\,.
}
Performing the same changes of coordinates \eqref{eq:trApp} as in the Maxwell case, and imposing the roots in $x=\pm 1$ by eliminating $c_3$ and $c_4$, we find the new form of the solution, namely
\ba\label{eq:fM Cmetric ModMax}
f_{\mbox{\tiny ModM}}&=&{\frac{r^2}{\ell^2}\Bigl(1-{\cal A}^2\ell^2\bigl[z^2{\cal A}^2-\frac{c_2}{2}\bigr]\Bigr)
-\frac{{\cal A}^2c_1}{6}r-\frac{c_2}{2}+\frac{c_1}{6r}+\frac{z^2}{r^2}}
\,,\nonumber\\
h_{\mbox{\tiny ModM}}&=&(1-x^2) \left(\mathcal{A}^2 z^2 \left(x^2+1\right)-\frac{\mathcal{A} c_1 x}{6}-\frac{c_2}{2}\right)\,, \label{eq:hM ModMax}\\
\Omega &=& 1+\mathcal{A}rx\,, \nonumber
\ea
and the solution reads
\ba\label{eq:line element Maxwell Cmetric ModMax}
\dd s^2&=&\frac{1}{\Omega^2}\Bigl(-\frac{f_{\mbox{\tiny ModM}} \dd t^2}{\omega^2}+\frac{\dd r^2}{f_{\mbox{\tiny ModM}}}+\frac{r^2 \dd x^2}{h_{\mbox{\tiny ModM}}}+\frac{r^2 h_{\mbox{\tiny ModM}} \dd\varphi^2}{K^2}\Bigr)\,,\nonumber\\
B&=&-\frac{e}{r}\exp(-\gamma)\frac{\dd t}{\omega}+\frac{x g \dd\varphi}{K}\,.
\ea
The solution reduces to that in Maxwell theory upon setting $\gamma=0$; note that the relative factors $\exp(-\gamma)$ in between the magnetic and electric parameters are slightly different in the vector potential $B$ than in the metric functions $F$ and $G$ (where they are entirely encoded via the parameter $z$).

\subsection{Thermodynamics}

 Following the Maxwell case, we can now find all thermodynamic charges, generalizing the results in \cite{Barrientos:2022bzm} to arbitrary parametrization $c_1$ and $c_2$. Solving the horizon equation $f_{\mbox{\tiny ModM}}(r_+)=0$ yields
\be\label{c1_eqMM}
c_1=\frac{r_+^2(3c_2-6{\cal A}^2z^2)-6z^2}{r_+}+\frac{6 r_+^3}{\ell^2({\cal A}^2 r_+^2-1)}\,. 
\ee 
We then find
\ba
M&=&\frac{c_1 \left(\mathcal{A}^2 l^2 \left(4 \mathcal{A}^2 z^2-c_2\right)-2\right)}{24 K\omega}\,,\nonumber\\
T&=&\frac{f'_{\mbox{\tiny ModM}}(r_+)}{4\pi \omega}=\frac{c_2 \ell^2 r_+^2 \left(\mathcal{A}^2 r_+^2-1\right)^2-2 z^2 \ell^2 \left(\mathcal{A}^2 r_+^2-1\right)^3+2 r_+^4 \left(\mathcal{A}^2 r_+^2-3\right)}{8 \pi  \ell^2 r_+^3 \omega  \left(\mathcal{A}^2 r_+^2-1\right)}\,,\nonumber\\
S&=&\frac{\pi r_+}{K(1-\mathcal{A}^2r_+^2)}\,,\quad 
\phi=-B \cdot \xi|_{H}=\frac{e \exp(-\gamma)}{\omega r_+}\,,\quad 
Q=\frac{1}{4\pi} \int_{S^2} *D=\frac{e}{K}\,,\nonumber\\
Q_m&=&\frac{1}{4\pi}\int_{S^2} F=\frac{g}{K}\,,\quad \phi_m=\frac{g \exp(-\gamma)}{\omega r_+}\,,\quad P=\frac{3}{8\pi \ell^2}\,,\\
\mu_+&=&\frac{-4 \mathcal{A}^2 z^2+c_2+2 K}{4K}\,,\quad 
\mu_-=-\frac{c_1 \mathcal{A}}{12K}\,.\nonumber
\ea
To calculate the electromagnetic charges we have employed the  generalized expressions for NLEs \eqref{QmQ}. 
We have also employed electromagnetic duality (valid for the ModMax theory) to calculate the magnetic potential $\phi_m$. {Note that many of the expressions (those coming from calculations using only the metric) agree with the Maxwell expressions \eqref{TDs_general} and \eqref{eq:mupp Maxwell}, upon replacing simply $e^2\to z^2$.
At the same time, the thermodynamic quantities derived from the vector potential are distinct from those of Maxwell electrodynamics by appearance of various extra factors of $\exp(-\gamma)$. }

 It is then easy to verify that we obey the following extended first law and Smarr relations \footnote{The Smarr relation again follows from Euler's theorem since $(M, S, P, Q, Q_m) \sim (L^1, L^2, L^{-2}, L^1, L^1)$ bearing in mind that $g \sim L^1$ and $\gamma \sim L^0$.}
\ba\label{eq:first law ModMax}
\delta M&=&T\delta S+\phi \delta Q+\phi_m \delta Q_m+\lambda_+ \delta\mu_+ + \lambda_- \delta\mu_- +V\delta P\,,\nonumber\\
M&=&2(TS-PV)+\phi Q+\phi_m Q_m\,,
\ea
provided we choose
\be\label{eq:omega ModMax generic solution}
\omega=\frac{1}{2}\sqrt{\left(4 \mathcal{A}^2 z^2-c_2\right) \left(2-4 \mathcal{A}^4 z^2 \ell^2+\mathcal{A}^2 c_2 \ell^2\right)}\,,
\ee
together with the following thermodynamic lengths:
\ba
\lambda_+&=&{\frac{c_2 \ell^2 r_+^2 \left(\mathcal{A}^2 r_+^2+1\right)-2 z^2 \ell^2 \left(\mathcal{A}^4 r_+^4+4 \mathcal{A}^2 r_+^2-1\right)+2 r_+^4}{2\ell^2 r_+ \omega \left(\mathcal{A}^2 r_+^2-1\right) \left(c_2-4 \mathcal{A}^2 z^2\right)}}\,, \label{eq:ModMax conjugate}\nonumber\\
\lambda_-&=&{\frac{\mathcal{A} \left(-4 \mathcal{A}^4 z^2 \ell^2 r_+^2+c_2 \ell^2 \left(\mathcal{A}^2 r_+^2-1\right)+4 \mathcal{A}^2 z^2 \ell^2+2 r_+^2\right)}{2 \omega\left(\mathcal{A}^2 r_+^2-1\right)}}\,,\\
V&=&\pi\frac{\mathcal{A}^4 \left(c_1 \ell^4 \left(c_2 r_+^2+4 z^2\right)+24 z^2 \ell^2 r_+^3\right)-\mathcal{A}^2 \ell^2 \left(c_1 c_2 \ell^2-4 c_1 r_+^2+12 c_2 r_+^3-24 z^2 r_+\right)-4 \mathcal{A}^6 c_1 z^2 \ell^4 r_+^2-24 r_+^3}{18 K \omega\left(\mathcal{A}^2 r_+^2-1\right)}\,,\nonumber\quad
\ea   
simply replacing $e^2\to z^2$ in the Maxwell formulas. 
Of course, one can still exploit the freedom in the choice of $c_2$ to further simplify the resultant thermodynamics. In particular, one can consider a parametrization where $\omega=1$.

We have also verified that the topological action renormalization yields the following result
\be 
F_T=\frac{S_T}{\beta}=F-2\mu_+ S_0 T=M-\phi Q-T(S+2\mu_+S_0)\,,
\ee 
similar to the Maxwell case.

%%%%%%%%%%%%%%%%%%%%%%%%%%%%%%
%%%%%%%%%%%%%%%%%%%%%%%%%%%%%%%%
\section{RegMax C-Metric} \label{regmax}

 Let us finally turn to the charged accelerating black holes in RegMax nonlinear electrodynamics discovered in  \cite{Kubiznak:2023emm}. We start from the $x-y$ form of the solution presented in Appendix~A of  \cite{Kubiznak:2023emm}, and proceed in a manner similar to the Maxwell case.

\subsection{Solution}

The field equations \eqref{NLEFE}
have the solution
\cite{Kubiznak:2023emm}
\ba\label{XYansatz}
\dd s^2&=&\frac{1}{H\!(x,y)^2}\Bigl(-F(y)\dd t^2+\frac{\dd y^2}{F(y)}+\frac{\dd x^2}{G(x)}+G(x)\dd\varphi^2\Bigr)\,, \nonumber\\
B&=&-\frac{\alpha e y}{\alpha+y\sqrt{|e|}}\dd t\,.
\ea
Here, similar to the Maxwell case  $H=x+y$ and
\ba
F&=&-4\alpha^4\log\Bigl(1+\frac{\sqrt{|e|}y}{\alpha}\Bigr)+\Bigl(\frac{c_1}{6}+\frac{4|e|^{3/2}\alpha}{3}\Bigr)y^3-\Bigl(\frac{c_2}{2}+2|e|\alpha^2\Bigr)y^2+\Bigl(c_3+4\sqrt{|e|}\alpha^3\Bigr)y-c_4+\frac{1}{\ell^2}\,,\nonumber\\
G&=&4\alpha^4\log\Bigl(1-\frac{\sqrt{|e|}x}{\alpha}\Bigr)+\Bigl(\frac{c_1}{6}+\frac{4|e|^{3/2}\alpha}{3}\Bigr)x^3+
\Bigl(\frac{c_2}{2}+2|e|\alpha^2\Bigr)x^2+\Bigl(c_3+4\sqrt{|e|}\alpha^3\Bigr)x+c_4\,.
\ea
Contrary to the form presented in \cite{Kubiznak:2023emm}, we have shifted the integration constants $\{c_1,c_2,c_3\}$, so that we recover the Maxwell solution \eqref{eq:Cmetric XY ansatz M} in the limit $\alpha\to \infty$.

Redefining coordinates and metric functions
as in \eqref{eq:trApp}
and \eqref{eq:trApp2} (and setting $f_M\to f, h_M\to h$)
yields the solution in the following form
\ba\label{Cansatz}
\dd s^2&=&\frac{1}{\Omega^2}\Bigl(-f\frac{\dd t^2}{\omega^2}+\frac{\dd r^2}{f}+r^2\Bigl[\frac{\dd x^2}{h}+ h \frac{\dd\varphi^2}{K^2}\Bigr]\Bigr)\,,\nonumber\\ 
B&=&-\frac{\alpha e}{\alpha r+\sqrt{|e|}}\frac{\dd t}{\omega}\,,
\ea
and the conformal factor given by 
\be\label{OmegaHl}
\Omega=1+{\cal A}rx\,. 
\ee

Imposing finally the requirement that the metric  function $h$ has roots $x=\pm 1$  (and restricting to the interval $x\in (-1,1)$--ensuring  a spherical horizon topology)  
fixes two of the integration constants to be 
\ba
c_3&=&\frac{2\alpha^4}{{\cal A}}\log\Bigl(\frac{\alpha+\sqrt{|e|}{\cal A}}{\alpha-\sqrt{|e|}{\cal A}}\Bigr)-4\sqrt{|e|}\alpha^3-\frac{4}{3}{\cal A}^2|e|^{3/2}\alpha-\frac{c_1}{6}{\cal A}^2\,,\label{eq:c3 RegMax}\\
c_4&=&-2\alpha^4\log\Bigl(\frac{\alpha^2-|e|{\cal A}^2}{\alpha^2}\Bigr)-2{\cal A}^2|e|\alpha^2-\frac{c_2}{2}{\cal A}^2\,,\label{eq:c4 RegMax}
\ea
upon which the metric functions $f$ and $h$ take the following form
\ba 
f_{\mbox{\tiny RegM}}&=&\frac{2r\alpha^4}{{\cal A}}\log\Bigl(\frac{\alpha+\sqrt{|e|}{\cal A}}{\alpha-\sqrt{|e|}{\cal A}}\Bigr)+2r^2\alpha^4\log\Bigl(\frac{\alpha^2-|e|{\cal A}^2}{\alpha^2}\Bigr)-4\alpha^4r^2\log\Bigl(1+\frac{\sqrt{|e|}}{r\alpha}\Bigr)\nonumber\\
&&\qquad +({\cal A}^2r^2-1)\Bigl(2|e|\alpha^2-\frac{4\alpha |e|^{3/2}}{3r}+\frac{c_2}{2}-\frac{c_1}{6r}\Bigr)+\frac{r^2}{\ell^2}\,,\nonumber\\
h_{\mbox{\tiny RegM}}&=&
\frac{2x\alpha^4}{{\cal A}^2}\log\Bigl(\frac{\alpha+\sqrt{|e|}{\cal A}}{\alpha-\sqrt{|e|}{\cal A}}\Bigr)-\frac{2\alpha^4}{{\cal A}^2}\log\Bigl(\frac{\alpha^2-|e|{\cal A}^2}{\alpha^2}\Bigr)+\frac{4\alpha^4}{{\cal A}^2}\log\Bigl(1-\frac{\sqrt{|e|}{\cal A}x}{\alpha}\Bigr)\nonumber\\
&&\qquad +(x^2-1)\Bigl(2|e|\alpha^2+\frac{4\alpha{\cal A} |e|^{3/2}x}{3}+\frac{c_2}{2}+\frac{{\cal A}c_1 x}{6}\Bigr)\,.
\ea 
In what follows we shall discuss properties of the solution in this general parametrization. Moreover, to shorten the (already very long) expressions, we shall (WLOG) concentrate on the case of positive charge, $e>0$.

%%%%%%%%%%%%%%%%%%%%%%%%%%%%%%%%%%
%%%%%%%%%%%%%%%%%%%%%%%%%%%%%%%%%
\subsection{Thermodynamic charges} \label{sec:td}

As in the Maxwell case, we can use the horizon equation 
\be 
f_{\mbox{\tiny RegM}}(r_+)=0\,
\ee 
to express one of the integration constants, say $c_1$, in terms of the horizon radius $r_+$, namely 
\ba \label{c1regmax}
c_1&=&\frac{6r_+}{{\cal A}^2r_+^2-1}\Bigl[\frac{2r_+\alpha^4}{{\cal A}}\log\Bigl(\frac{\alpha+\sqrt{e}{\cal A}}{\alpha-\sqrt{e}{\cal A}}\Bigr)+2r_+^2\alpha^4\log\Bigl(\frac{\alpha^2-e{\cal A}^2}{\alpha^2}\Bigr)-4\alpha^4r_+^2\log\Bigl(1+\frac{\sqrt{e}}{r_+\alpha}\Bigr)\nonumber\\
&&\qquad +({\cal A}^2r_+^2-1)\Bigl(2e\alpha^2-\frac{4\alpha e^{3/2}}{3r_+}+\frac{c_2}{2}\Bigr)+\frac{r_+^2}{\ell^2}\Bigr]\,.
\ea 
Since inserting this expression
into the metric
further complicates the thermodynamic expressions that follow, wherever possible we shall keep $c_1$ and understand it as a shorthand for the right-hand side of \eqref{c1regmax}.

Computation of the thermodynamic charges associated with the accelerating  RegMax black holes proceeds in a manner similar to the Maxwell case.  The mass is
\ba
M&=&\frac{1}{24 \mathcal{A} K \omega  \left(\alpha ^2-\mathcal{A}^2 e\right)^2}\Bigl[ 12 \alpha ^4 \ell^2 \left(\alpha ^2-\mathcal{A}^2 e\right) \left(\mathcal{A}^2 e \left(c_2+4 \alpha ^2 e\right)-\alpha ^2 c_2\right) \tanh ^{-1}\Bigl(\frac{\mathcal{A} \sqrt{e}}{\alpha }\Bigr) \quad\nonumber\\
&&\quad - \mathcal{A}^7 \ell^2 e^2 \left(c_1+8 \alpha  e^{3/2}\right) \left(c_2+4 \alpha ^2 e\right)-2 \alpha ^4 \mathcal{A} \left(c_1-6 \alpha ^3 c_2 \ell^2 \sqrt{e}\right)- \alpha ^2 \mathcal{A}^3c_1 \left(\alpha ^2 c_2 \ell^2-4 e\right)\\
&&\quad  - 4\alpha^3 \mathcal{A}^3 e^{3/2} \left(5 \alpha ^2 c_2 \ell^2+12 \alpha ^4 \ell^2 e-8 e\right) +2 \mathcal{A}^5 e \left(c_1+8 \alpha  e^{3/2}\right) \left(\alpha ^2 c_2 \ell^2+e \left(2 \alpha ^4 \ell^2-1\right)\right) \Bigr]\,.\quad\nonumber
\ea
The calculation of the temperature and entropy is standard and yields:
\ba 
T&=&\frac{f_{\mbox{\tiny RegM}}'(r_+)}{4\pi\omega}=
-\frac{c_1}{24\pi\omega r_+^2}-\frac{ \alpha  e^{3/2}}{3\pi\omega r_+^2}+\frac{ \alpha ^4 r_+ \sqrt{e}}{\pi\omega(\sqrt{e}+\alpha  r_+)}+\frac{ r_+}{2\pi\omega \ell^2}-\frac{\mathcal{A}^2 \left(c_1-6 r_+ \left(c_2+4 \alpha ^2 e\right)+8 \alpha  e^{3/2}\right)}{24 \pi  \omega }\nonumber\\
&&\qquad +\frac{\alpha ^4 }{2 \pi  \mathcal{A} \omega }\left(2 \mathcal{A} r_+ \log \left(1-\frac{\mathcal{A}^2 e}{\alpha ^2}\right)+\log \left(\frac{\alpha +\mathcal{A} \sqrt{e}}{\alpha -\mathcal{A} \sqrt{e}}\right)-4 \mathcal{A} r_+ \log \left(\frac{\sqrt{e}}{\alpha  r_+}+1\right)\right)\,,\\
S&=&\frac{\mbox{Area}}{4}=\frac{\pi r_+^2}{K(1-\mathcal{A}^2 r_+^2)}\,.
\ea

%}
%\tcr{Yes, the blue one should be correct.}
The  electric charge now reads \be\label{eq:chargedef}
Q=\frac{1}{4\pi}\int_{S^2} * D=\frac{\alpha^2 e}{K(\alpha^2-{\cal A}^2e)}\,,
\ee
using the formula \eqref{QmQ}.
It is  tempting to define the thermodynamic electrostatic potential according to  
\be \label{phi0ModMax}
\phi_0=\left.-\xi \cdot B\right|_{r=r_+}=\frac{\alpha e}{\omega(\alpha r_++\sqrt{e})}
\ee
but  this is not consistent  with the Euclidean action calculation. 
%does not give rise to consistent thermodynamics. 
We instead use the Hawking-Ross boundary prescription \cite{Hawking:1995ap}, which in the case of nonlinear electrodynamics generalizes to \cite{Barrientos:2022bzm}
\be 
\phi=\frac{1}{4\pi \beta Q}\int_{\partial M} \dd^3x\sqrt{h} n_\mu D^{\mu\nu} B_\nu\,,
\ee 
where the vector potential $B$ is considered in a gauge in which it vanishes on the black hole horizon, i.e. $B\to B+\phi_0 \dd t$. This  yields
\be\label{phiRM}
\phi=\phi_0+\frac{K {\cal A}^2 \sqrt{e} Q}{\omega \alpha}=\frac{\alpha e}{\omega(\alpha r_++\sqrt{e})}-
\frac{\alpha e}{\omega(\alpha r_{\cal A}+\sqrt{e})}\,,
\ee
where we have defined $r_{\cal A}\equiv-\alpha/({\cal A}^2 \sqrt{e})$. Formally, such a potential "looks like" a difference between potentials evaluated at $r_+$ and $r_A$.
As we shall see, it is this potential that also emerges from the topological action renormalization.

The string tensions now read 
\be
\mu^{\pm}\equiv\frac{\delta_{\pm}}{8\pi}\equiv\frac{1}{4}\Bigl(1-\frac{{h_{\mbox{\tiny RegM}}}}{K(1-x^2)}\Bigr)\Big|_{x=\pm 1}\equiv \frac{1}{4}\left(1-\frac{\Xi\pm\sigma}{K}\right)\,,
\ee
where 
\eq{
\Xi=&\frac{\alpha ^2 c_2-\mathcal{A}^2 e \left(c_2+4 \alpha ^2 e\right)}{2 \mathcal{A}^2 e-2 \alpha ^2}\,,\nonumber\\
\sigma=&\frac{\mathcal{A}^5 (-e) \left(c_1+8 \alpha  e^{3/2}\right)+\alpha ^2 \mathcal{A}^3 \left(c_1+8 \alpha  e^{3/2}\right)+12 \alpha ^4 \left(\alpha ^2-\mathcal{A}^2 e\right) \tanh ^{-1}\left(\frac{\mathcal{A} \sqrt{e}}{\alpha }\right)-12 \alpha ^5 \mathcal{A} \sqrt{e}}{6 \mathcal{A}^4 e-6 \alpha ^2 \mathcal{A}^2}\,.
}
From this we obtain 
\eq{
\mu_+\equiv & \mu^++\mu^-=\frac{1}{2}-\frac{\Xi}{2K}=\frac{\alpha ^2 \left(-4 \mathcal{A}^2 e^2+c_2+2 K\right)-\mathcal{A}^2 e (c_2+2 K)}{4 K \left(\alpha ^2-\mathcal{A}^2 e\right)}\,,\nonumber\\
\mu_-\equiv & \mu^--\mu^+=\frac{\sigma}{2K}=\frac{\mathcal{A}^5 e \left(c_1+8 \alpha  e^{3/2}\right)-\alpha ^2 \mathcal{A}^3 \left(c_1+8 \alpha  e^{3/2}\right)-12 \alpha ^4 \left(\alpha ^2-\mathcal{A}^2 e\right) \tanh ^{-1}\left(\frac{\mathcal{A} \sqrt{e}}{\alpha }\right)+12 \alpha ^5 \mathcal{A} \sqrt{e}}{12 \mathcal{A}^2 K \left(\alpha^2-\mathcal{A}^2 e\right)}\,.
}
\iffalse
\tcr{\bf For the new metric function $h$, I obtained these tensions (with the right Maxwell limit):}
\tcr{
\be
\mu_+\equiv & \mu^++\mu^-=\frac{\frac{4 \alpha ^2 \mathcal{A}^2 e^2}{\mathcal{A}^2 e-\alpha ^2}+c_2+2 K}{4 K}\,,
\ee
\be
\mu_-\equiv & \mu^--\mu^+=-\frac{\mathcal{A}^5 e \left(c_1+8 \alpha  e^{3/2}\right)-\alpha ^2 \mathcal{A}^3 \left(c_1+8 \alpha  e^{3/2}\right)+3 \alpha ^4 \left(\alpha ^2-\mathcal{A}^2 e\right) \log \left(\frac{\left(\alpha -\mathcal{A} \sqrt{e}\right)^2}{\left(\alpha +\mathcal{A} \sqrt{e}\right)^2}\right)+12 \alpha ^5 \mathcal{A} \sqrt{e}}{12 \mathcal{A}^2 K \left(\mathcal{A}^2 e-\alpha ^2\right)}\,.
\ee}
\fi
Of course, all of the above thermodynamic quantities reduce to the Maxwell case in the limit of large $\alpha$.

Since $\alpha$ is an additional dimensional parameter, we now seek the following extended first law and Smarr relations \footnote{Again, Eq.~\eqref{Smarr_RegMax} can be derived using Euler's theorem, as $(M, S, P, Q, \alpha) \sim (L^1, L^2, L^{-2}, L^1,L^{-1/2})$.}:
\ba 
\delta M&=&T \delta S+\phi \delta Q+\lambda_+\delta \mu_++\lambda_-\delta \mu_-+V \delta P+\Pi_\alpha \delta \alpha\,,\\
M&=&2(TS-PV)+\phi Q-\frac{1}{2}\Pi_\alpha \alpha\,, \label{Smarr_RegMax}
\ea 
where we have also included the standard $P-V$ term, and $\lambda_+, \lambda_-$, and $\Pi_\alpha$ are the corresponding conjugate quantities.

Even in this (very algebraically nontrivial) case, we have verified that the topological action renormalization yields the result 
\eqref{FFT}, namely
\be 
F_T+2\mu_+S_0T=F=M-TS-\phi Q\,.
\ee 
Note that in here, we have the electrostatic potential $\phi$, \eqref{phiRM}, obtained from the Hawking-Ross prescription, rather than the naive $\phi_0$, \eqref{phi0ModMax}. In other words, the topological renormalization gives rise to an upgraded electrostatic potential, similar to what happens in the special case studied in the Appendix. The origin of this modified potential remains to be better understood.

To complete the thermodynamic description, the last step would be to find an expression for $\omega$, generalizing \eqref{eq:omega Maxwell generic solution 1}. 
Unfortunately, due to the complexity of the above expressions, we were unable to solve the corresponding partial differential equation. In fact, even the first correction $O(1/\alpha)$  seems hard to find. For this reason, we have to leave $\omega$ unspecified at the moment, leaving the verification of the first law for future work.

\section{Conclusions} \label{sec:concl}

We have revisited  C-metric spacetimes describing charged and (slowly) accelerated black holes in various theories of (nonlinear) electrodynamics in asymptotically AdS spacetimes and revised their thermodynamics. Starting from a general parametrization of the metric, we have shown that it is possible to formulate thermodynamic laws with an additional parameter. Exploiting the corresponding freedom, we have demonstrated that one can, for example, eliminate the nontrivial normalization of the boost Killing vector, which was a peculiar feature of previous studies of thermodynamics of these black holes. Interestingly, the process of fixing the extra parameter is not just a "gauge choice"--as, for example,  explicitly demonstrated in Sec.~\ref{2C3}, where it partially fixes the distribution of cosmic strings--and consequently may have physical implications.   
In particular, it suggests that a choice of the parameter $c_2$ may affect the choice of the thermodynamic ensemble (and consequently may affect the phase behavior of the black hole thermodynamic system).

 We have also calculated the Euclidean action for the corresponding spacetimes using (i) the standard holographic renormalization procedure, and (ii) the topological renormalization. We have shown that in the presence of nontrivial overall conical deficit $\mu_+$, the two are not equal and are related by the  formula
\be \label{master}
F_T=F+2\pi \ell^2 \mu_+ T=M-\phi Q-T\tilde S\,,\quad \tilde S=S-2\pi \ell^2 \mu_+\,, 
\ee 
in  the Maxwell, ModMax, and RegMax cases.
This result raises various questions and is perhaps the most interesting finding of our study. For example, does this mean that there are additional boundary terms related to cosmic strings (as encoded in the topological action renormalization calculation), similar to what happens in the three-dimensional C-metric spacetimes \cite{Arenas-Henriquez:2023hur, Cisterna:2023qhh, Arenas-Henriquez:2024ypo}? Does it mean that, due to the presence of cosmic strings,  one could also consider a modified entropy $\tilde S$ above, instead of the Bekenstein area law? Does it imply  that accelerated black hole spacetimes can be assigned a noninteger (generalized) Euler characteristic, as proposed in \cite{Cassani:2021dwa}?

 We have also extended previous studies of charged accelerated black holes to ModMax and RegMax theories of nonlinear electrodynamics. While the former seems to be a straightforward generalization of the Maxwell case (and was already studied in a standard parametrization in \cite{Barrientos:2022bzm}), the latter is completely new and much more algebraically involved. That is why, while we were able to construct all thermodynamic quantities in this case, we have yet to obtain the corresponding normalization of the boost Killing vector and thence verify the first law. A new very interesting feature of RegMax accelerated black holes (absent in the standard Maxwell and ModMax cases) is the modification of the electrostatic potential. This can be calculated by the Hawking-Ross prescription and agrees with the topological action renormalization; formally such a potential can be written as a difference of potentials on two different radii (one of which is negative). This modification as well appears in the Maxwell case upon employing the special parametrization (see Appendix) and in both cases remains to be understood (perhaps using the approach of \cite{Elgood:2020svt}). Its structure reminds one of a modified angular velocity for, for example, Kerr-AdS spacetimes where a difference between the horizon quantity and the nontrivial rotation at infinity has to be considered \cite{Gibbons:2004ai}. Understanding the origin of this modification of the potential may be crucial for completing the first law for accelerating RegMax black holes.

In all three cases studied, the expression \eqref{master} for the free energy remains valid. This suggests (as may be expected) that it is a universal feature of cosmic string spacetimes, rather than a consequence of their specific  electromagnetic field content. The ultimate understanding of this formula is left for future studies.

\section*{ACKNOWLEDGEMENTS}

We are grateful to Rodrigo Olea for useful discussions.
 D.K. is grateful for support from
GAČR 23-07457S grant of the Czech Science Foundation
and the Charles University Research Center Grant No.
UNCE24/SCI/016. 
This work was supported in part by the Natural
Sciences and Engineering Research Council of Canada.
Research at Perimeter Institute is supported in part by the Government of Canada through the Department of Innovation, Science and Economic Development and by the Province of Ontario through the Ministry of Colleges and Universities.

\section*{DATA AVAILABILITY}
No data were created or analyzed in this study.

\appendix

\section{MAXWELL CASE: SPECIAL PARAMETRIZATION}\label{sec:appA}
\begin{figure}[H]
\begin{center}
    \includegraphics[scale=0.67]{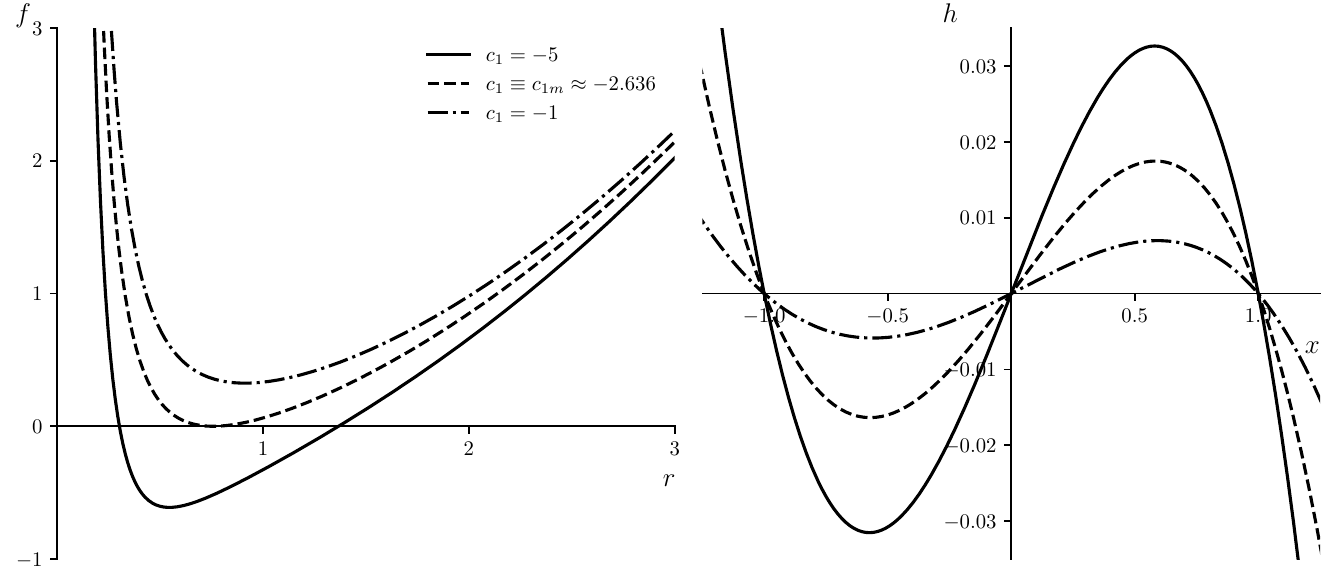}
	\caption{{\textbf{Black hole horizons.} Upon parametrization \eqref{AdSpar} (with $\mathcal{A}=0.1, e=0.5, \ell=2$), the black hole can have two, one, or no horizons. For $c_1<0$, it holds that $h(x)>0$ on $x\in(0,1)$.}}
\label{fig:fh_2a2e2}
\end{center}
\end{figure}

In the general case, the Maxwell metric function $f_M$ contains the AdS term $\frac{r^2}{\ell^2}$ which is, however, disrupted by another term with $r^2$ dependence. In order to get rid of this term and obtain a purely AdS asymptotic, one can choose $c_2$ to be
\be
c_2=2\mathcal{A}^2 e^2. \label{AdSpar}
\ee
With this choice, our metric functions take on the following form:
\be
f_M=-\frac{1}{6} \mathcal{A}^2 \left(c_1 r+6 e^2\right)+\frac{c_1}{6 r}+\frac{e^2}{r^2}+\frac{r^2}{\ell^2},
\ee
\be
h_M=-\frac{1}{6} \mathcal{A} (x-1) x (x+1) \left(6 \mathcal{A} e^2 x-c_1\right).
\ee
The metric function $f_M$ still describes a black hole whose horizon structure is depicted in Fig.~\ref{fig:fh_2a2e2}.
Note, however, that in this case,   $h_M$ has   gained a new root at $x=0$; it is positive on the interval $(0,1)$ and negative on $(-1;0)$ (as can be seen in Fig.~\ref{fig:fh_2a2e2}). Therefore, to calculate any thermodynamic quantities, one must restrict integration over $x$ to the positive interval. The parametrization \eqref{AdSpar} is thus fundamentally different from other parametrizations presented in Sec.~\ref{sec:maxwell}.

In this case we find the following thermodynamic quantities:
\ba
T&=&\frac{e^2 \ell^2 \left(\mathcal{A}^2 r_+^2-1\right)^2+r_+^4 \left(\mathcal{A}^2 r_+^2-3\right)}{4 \pi  \ell^2 r_+^3 \omega  \left(\mathcal{A}^2 r_+^2-1\right)}\,,\quad S=\frac{\pi  r_+^2}{2 K (\mathcal{A} r_++1)}\,,\nonumber\\ 
Q&=&\frac{e}{2 K}\,,\quad \phi=\frac{e}{\omega r_+}-\frac{e}{\omega r_{\cal A}}\,,\quad r_{\cal A}\equiv -\frac{2}{\cal A}\,,\nonumber\\
M&=&\frac{\left(12 \mathcal{A} e^2-c_1\right) \left(\mathcal{A}^3 \ell^2 \left(c_1-4 \mathcal{A} e^2\right)+16\right)}{384 K \omega }\,,\nonumber\\
\mu_+&=&\frac{\mathcal{A}(c_1-4{\cal A}e^2)}{16K}+\frac{1}{2}\,,\quad \mu_-=\frac{\mathcal{A}}{48K}(12e^2{\cal A}-c_1)\,.
\ea 
 Similar to the RegMax case in the main text, the modified potential $\phi$ (different from $\phi$ in \eqref{TDs_general}) can be calculated using the Hawking-Ross prescription, or using the (topological) action calculation. We find in particular that
\be 
F=M-TS-\phi Q=F_T+2\mu_+S_0 T 
\ee 
with  $\phi$ as written above. The origin of the modification of $\phi$ in this (and the RegMax) case remains to be understood. 

Let us first comment on the uncharged case, $e=0$ (assuming $c_1<0$). In this case we find that the variations $\delta \mu_+$ and $\delta \mu_-$ are no longer independent; we have 
\be \label{mu3mu}
\delta \mu\equiv\delta \mu_+=-3\delta \mu_-\,.
\ee
Then we find that the  following thermodynamic laws
\ba 
\delta M&=&T \delta S+\lambda \delta \mu+V \delta P\,,\nonumber\\ \label{flspecial}
M&=&2TS-2PV\, \label{smarrspecial}
\ea 
are satisfied provided in addition to the above, we set 
\ba 
\omega&=&\frac{r_+ \sqrt{\mathcal{A}r_+ (8-3 \mathcal{A}^3 r_+^3-8 \mathcal{A}^2 r_+^2)}}{\ell(1-\mathcal{A}^2 r_+^2)}\,,\quad 
V=\frac{\pi  r_+^3 \left(11 {\cal A}^3 r_+^3-24 {\cal A} r_++16\right)}{24 K \omega \left(1-{\cal A}^2 r_+^2\right)^2}\,,\nonumber\\
\lambda&=&\frac{4-12{\cal A}r_++4 {\cal A}^2r_+^2+7 {\cal A}^3r_+^3}{12 {\cal A}\omega(1-{\cal A}^2r_+^2)}\,.
\ea 

{More generally, for nontrivial $e$, $\delta \mu_+$, and $\delta \mu_-$ become independent, and  we find the following $\omega$:
\be 
\omega=\frac{\sqrt{\mathcal{A}(r_+^4+e^2\ell^2-{\cal A}^2e^2r_+^2\ell^2)\bigl[{\cal A}^3e^2\ell^2(2{\cal A}^3r_+^3+3{\cal A}^2r_+^2-2{\cal A}r_+-3)-3{\cal A}^3 r_+^4-8{\cal A}^2r_+^3+8r_+\bigr]}}{r_+\ell(1-{\cal A}^2r_+^2)}\,,
\ee 
which reduces to the previous expression upon setting $e\to 0$. This is then accompanied by generalized $\lambda_\pm$ and $V$, namely
\ba
V&=&\frac{\pi}{24 K \omega r_+^2 \left({\cal A}^2 r_+^2-1\right)^2}
\Bigl[4 {\cal A}^9 e^4 \ell^4 r_+^6+8 {\cal A}^8 e^4 \ell^4 r_+^5-5 {\cal A}^7 e^4 \ell^4 r_+^4-8 {\cal A}^6 e^2 \ell^2 r_+^3 \left(2 e^2 \ell^2+r_+^4\right)
\nonumber\\
&&+8 {\cal A}^4 e^2 \ell^2 r_+ \left(e^2 \ell^2+r_+^4\right)+{\cal A}^3 \left(3 e^4 \ell^4+6 e^2 \ell^2 r_+^4+11 r_+^8\right)-24 {\cal A} r_+^6+16 r_+^5-2 {\cal A}^5 \left(e^4 \ell^4 r_+^2+3 e^2 \ell^2 r_+^6\right)\Bigr]\,,\nonumber\\
\lambda_-&=&\frac{1- {\cal A}^4 e^2 \ell^2}{\omega{\cal A}}-\frac{15 {\cal A}^2 e^2 \ell^2}{16r_+}-\frac{1}{32\omega({\cal A}^2 r_++{\cal A})}-\frac{ e^2 \ell^2 r_+ \left({\cal A}^2 r_+^2-1\right) \left({\cal A}^4 e^2 \ell^2-4\right)}{4\omega\left( e^2 \ell^2 \left({\cal A}^2 r_+^2-1\right)-r_+^4\right)}+\frac{15}{32\omega{\cal A} \left({\cal A} r_+-1\right)}+\frac{7 r_+}{16\omega}\,,\nonumber\\
\lambda_+&=&-\frac{ {\cal A}^3 e^2 \ell^2}{4\omega}-\frac{ {\cal A}^2 e^2 \ell^2}{16\omega r_+}+\frac{17}{32\omega({\cal A}^2 r_++{\cal A})}-\frac{ e^2 \ell^2 r_+ \left({\cal A}^2 r_+^2-1\right) \left({\cal A}^4 e^2 \ell^2-4\right)}{4\omega\left(e^2 \ell^2 \left({\cal A}^2 r_+^2-1\right)-r_+^4\right)}+\frac{1}{32\omega{\cal A} \left({\cal A} r_+-1\right)}-\frac{7 r_+}{16\omega}\,.
\ea
While these expressions are not very illuminating, one can check that they satisfy the extended first law and Smarr relations \eqref{eq:first law Maxwell1} and \eqref{eq:first law Maxwell1b}.

%\bibliography{RegMaxCmetric}% Produces the bibliography via BibTeX.


\end{document}